\documentclass[12pt]{article}

\newcommand{\eps}{\epsilon}
\newcommand{\veps}{\varepsilon}
\newcommand{\nth}{^{\mbox{\scriptsize{th}}}}
\newcommand{\ii}{\mathrm{i}}
\newcommand{\wtild}{\widetilde}
\newcommand{\vphi}{\varphi}

\usepackage{amssymb}
\usepackage{amsmath}

\newtheorem{theorem}{Theorem}[section]

\newtheorem{remark}{Remark}[section]

\newtheorem{fact}{Fact}[section]

\newtheorem{fpf}{Fixed point factorisation}

\begin{document}
\centerline{{\LARGE Fixed points and fusion rings. Part 1}}

\bigskip\centerline{Elaine Beltaos}

\centerline{{\it Department of Mathematics and Statistics, Grant MacEwan University}}

\centerline{{\it 10700 - 104 Ave, Edmonton, AB CANADA T5J 4S2}}

\bigskip

\centerline{ beltaose@macewan.ca}

\bigskip

\noindent{\bf Abstract.} In the first of this two-part series, we find `fixed point factorisation' formulas, towards an understanding of the fusion ring of WZW models.  Fixed-point factorisation refers to the simplifications in the data of a CFT involving primary fields fixed by simple-currents.  Until now, it has been worked out only for SU($n$), where it has developed into a powerful tool for understanding the fusion rings of WZW models of CFT --- e.g. it has lead to closed formulas for NIM-reps and D-brane charges and charge-groups.  In this paper, we generalise these formulas to the other classical algebras, laying the groundwork for future applications to fusion rings (Part 2). We also discuss connections with the twining characters of Fuchs-Schellekens-Schweigert.

\bigskip


\section{Introduction}\label{sIntro}
\setcounter{equation}{0}

The primary fields of a WZW model correspond to fixed level $k$ highest weight representations $\lambda$ of the underlying affine algebra.  Their characters are holomorphic functions of a complex number $\tau \in \mathbb{H}$, the upper half plane of complex numbers with positive imaginary part; more precisely, \begin{equation}
\chi_\lambda(\tau) = q^{-c/24} \mbox{Tr}_\lambda q^{L_0} \ ,
\end{equation} where $q = e^{2\pi \ii \tau}$, $c$ is the central charge, and $L_0$ is the energy operator.  As with all RCFTs, they satisfy a modularity property \begin{eqnarray}
\chi_\lambda(-1/\tau) &=& \sum_\mu S_{\lambda\mu} \chi_\mu(\tau) \ ; \label{charmodS} \\
\chi_\lambda(1+\tau) &=& \sum_\mu T_{\lambda\mu} \chi_\mu(\tau) \ , \label{charmodT}
\end{eqnarray} where the sum is over all primaries $\mu$, and $S_{\lambda\mu}$, $T_{\lambda\mu}$ are complex numbers.  The matrices $S$ and $T$ defined by \eqref{charmodS}, \eqref{charmodT} resp. generate a representation of the modular group SL$_2(\mathbb{Z}) = \left\{ \left( \begin{array}{cc} a & b \\ c & d \end{array} \right) \ | \ a,b,c,d \in \mathbb{Z}, \det A = 1\right\}$; they are called \emph{modular data}, and they satisfy many properties which we will discuss in Section \ref{sBackground}.  Modular data also occur in various other contexts in physics and mathematics (e.g. finite groups, VOAs, subfactors --- see \cite{moddata} for more).

The $S$-matrix is the more important of the two, and is a fundamental quantity of an RCFT, as the matrix governing the modular transformation $\tau \mapsto -1/\tau$ of the RCFT characters, and through Verlinde's formula (see \eqref{Verlinde}), expressing the fusion coefficients of the fusion ring.  In this paper, we are interested in simplifying the entries $S_{\lambda\vphi}$ of the $S$-matrix corresponding to an affine Kac-Moody algebra where $\vphi$ is a fixed point --- by `fixed point', we mean a primary fixed by nontrivial simple-current symmetries.  For the WZW models, simple-current symmetries correspond directly with the centre of the Lie group and to symmetries of the extended Coxeter-Dynkin diagram \cite{Fuchs}.\footnote{With the exception of $E_8^{(1)}$ at level 2.} Fixed points can complicate many phenomena --- e.g. calculating NIM-rep coefficients, or see the modular invariant partition function classification of \cite{SU3, Revisited} --- so it is important to have tools to handle these difficulties.  The $S$-matrix simplification we address in this paper is referred to as `fixed point factorisation' (see Theorem 3.1 for details).  A simplification to this task is provided by the observation that the ratio $S_{\lambda\mu}/S_{0 \mu}$ is a polynomial in ratios $S_{\Lambda_n \, \mu}/S_{0\mu}$ , where $\Lambda_n$ are the fundamental representations and $0$ denotes the vacuum primary (this will be explained more precisely in Section \ref{sBackground}).  Thus we need only find an explicit fixed point factorisation at entries $S_{\Lambda_n  , \vphi}$ to establish that one exists.

Fixed point factorisation was first found to exist, for SU($n$), by Gannon-Walton \cite{FPF} --- they found that the $S$-matrix entries involving fixed points factored into $S$-matrix entries of SU($n/d$), where $d$ is a proper divisor of $n$, and $n/d$ is the order of the simple-current fixing $\vphi$.  We give a brief review of this case in Section \ref{ssA}.  Gaberdiel-Gannon later used fixed point factorisation to find NIM-rep coefficients (a NIM-rep is a nonnegative integer matrix representation of a fusion ring --- see e.g. \cite{moddata, GNIM} for an introduction to NIM-reps; a Lie-theoretic interpretation for the WZW models was given in \cite{GG}) --- and their D-brane charges on non-simply connected Lie groups SU($n$)/$\mathbb{Z}_d$, where $d$ is a proper divisor of $n$ and the $\mathbb{Z}_d$ are subgroups of the group of simple-currents of SU($n$)\cite{GG2, GG1}.  One consequence of this is a beautiful and unexpected relation between string theories on non-simply connected Lie groups and simply connected groups of smaller rank.

Our fixed point factorisation formulas, given in Section \ref{sfpf}, address the important first step of providing the tool to determine NIM-reps and their D-brane charges for all WZW models --- these will follow in \cite{BGW}.  Current work in this direction has already yielded surprisingly simple expressions for the associated NIM-rep coefficients.  Just as the partition function of a torus yields a modular invariant, the partition function associated to a cylinder yields a NIM-rep (though not every NIM-rep is a cylindrical partition function for a consistent CFT --- e.g. the tadpole NIM-reps occurring for $A_1^{(1)}$ at odd level \cite{DiFZ}).  The NIM-rep coefficients satisfy a Verlinde-like formula; however an explicit proof that this formula yields nonnegative integers (for simple-current invariants) has not yet been found in the literature.  Our expressions could provide the groundwork for such a proof of integrality.\footnote{As we have explicit formulas only at the fundamental weights, positivity of all NIM-rep coefficients would not follow from positivity of those involving fundamental weights.}

As mentioned above, the SU($n$) fixed point factorisation formula involved factors of type SU($n/d$).  It was not clear a priori whether a fixed point factorisation would exist in other cases, and if so, which algebras should play the role of the smaller-rank algebras, or what the formulas would look like. However, the answer that has emerged yields a remarkable twist --- not only are the smaller-rank algebras not of the same family as the original algebra (indeed, nontwisted algebras can yield twisted algebras and vice-versa), but they are precisely the orbit Lie algebras of Fuchs-Schellekens-Schweigert \cite{FSS, FRS}!\footnote{This was pointed out to the author by Terry Gannon.}  Given a simple or affine Lie algebra $\mathfrak{g}$, its orbit Lie algebra $\breve{\mathfrak{g}}$ is obtained from $\mathfrak{g}$ through a diagram-folding, or matrix-folding technique.  What they found was that the twining characters of $\mathfrak{g}$  could be expressed in terms of true characters of $\breve{\mathfrak{g}}$ (twining characters are characters that have been `twisted' by an automorphism of $\mathfrak{g}$).  We note that orbit Lie algebras have appeared before in a related context to ours.  When \emph{both} primaries involved are fixed points, the matrices $S^J$ of \cite{FSS2}, giving the $S$-matrix for the simple-current extended theory, can be identified with the $S$-matrix for the orbit Lie algebra associated with the simple-current $J$.

We remark that the exceptional cases --- namely $E_6^{(1)}$ and $E_7^{(1)}$ --- are yet to be worked out.  Their Weyl groups are irregular, but preliminary work on this, with the aid of Maple, suggests that their orbit Lie algebras ($G_2^{(1)}$ and $F_4^{(1)}$ resp.) again should be the smaller-rank fixed point factorisation algebras.  With some further techniques, we expect that explicit formulas should also be within reach.

\paragraph{Notation.} We use the notation of \cite{Kac} for the affine algebras: by $X_{r}^{(i)}$, where $X \in \{A, B, C, D, E, $ $F, G\}$ and $i \in \{1,2\}$, we mean the affine algebra with underlying rank-$r$ simple finite dimensional algebra $X_r$, twisted by an automorphism of order $i$.  We identify an $X_r$ representation with its Dynkin labels: $\lambda = (\lambda_1, \dots , \lambda_r)$, and similarly, for $X_r^{(i)}$, $\lambda=(\lambda_0; \dots , \lambda_r)$.  We denote the $n\nth$ fundamental weight by $\Lambda_n$; that is, $\Lambda_n = (0, \dots ,1, \dots 0)$, where the `1' is in the $n\nth$ position.

We let $P_+^k(X_r^{(i)})$ denote the set of level $k$ integrable highest weights for $X_r^{(i)}$ (the primaries); that is, \begin{equation} P_+^k(X_r^{(i)}) = \left\{ (\lambda_0; \dots , \lambda_r) \in \mathbb{N}^{r+1} \ : \ \sum_{\ell=0}^r a_\ell^\vee \lambda_\ell = k \right\} \ . \end{equation} The $a^\vee_\ell$ are the dual Coxeter labels of $X_r^{(i)}$.  For example, for $X=A,C$, and $i = 1$, $a_\ell^\vee = 1$ for all $0 \leq \ell \leq r$.

Throughout this paper, we let $\mathbb{N}$ be the set of nonnegative integers, and $^*$ denote complex conjugation.


\section{WZW modular data and simple-currents}\label{sBackground}
\setcounter{equation}{0}

In this section, we review the necessary information about WZW data and affine algebras.  Most of these properties hold for general RCFT as well.  In Section \ref{sApp}, we give the relevant information for the specific algebras involved in the fixed point factorisation formulas.  We assume general knowledge of Lie algebras and their representations --- some references are \cite{Kac, FH, FS}.

\subsection{Review and definitions}\label{subsec:SimpCur}

As we mentioned in the introduction, the matrices $S$ and $T$ defined in \eqref{charmodS} and \eqref{charmodT} generate a representation of SL$_2(\mathbb{Z})$.  More precisely, the representation is $\displaystyle{ \left(\begin{array}{lr} 0& -1 \\ 1&0\end{array}\right) \mapsto S }$ and $\displaystyle{ \left(\begin{array}{lr} 1&1 \\ 0&1\end{array}\right) \mapsto T}$.  The $S$-matrix is our main interest.  It is unitary and symmetric, i.e. $SS^* = I$, although we are also interested in the twisted $A$-algebras, for which symmetry fails --- we address the specifics of these cases in Section \ref{sApp}.  The matrix $S^2 =: C$ is a permutation matrix called charge-conjugation; it associates a $\mathcal{V}$-module to its dual (where $\mathcal{V}$ is the vertex operator algebra of holomorphic quantum fields). The $S$-matrix satisfies the following symmetry with respect to charge-conjugation: \begin{equation}\label{moddatsymmetries2} S_{C\lambda, \mu} = S_{\lambda,C\mu} = S^*_{\lambda\mu} \ . \end{equation}

The WZW models are unitary RCFTs, which means that we also have \begin{equation}\label{unitaryvacuum} S_{0\mu} \geq S_{00} > 0 \end{equation} for all primaries $\mu$, where $0$ denotes the vacuum.  Equality occurs for primaries called simple-currents (defined below); they also correspond to permutations of the set $P_+^k(X_r^{(i)})$ of primaries.  The most important property of the $S$-matrix is that the numbers $N_{\lambda\mu}^\nu$ defined by Verlinde's formula
\begin{equation}\label{Verlinde} N_{\lambda\mu}^\nu = \sum_{\alpha} \frac{S_{\lambda\alpha}S_{\mu\alpha}S^*_{\nu\alpha}}{S_{0\alpha}}
\end{equation} are nonnegative integers.  These numbers are called \emph{fusion coefficients}, and are structure constants for a commutative associative ring called the \emph{fusion ring}.  One consequence of the integrality of the fusion coefficients is a powerful Galois symmetry of the $S$-matrix (see \cite{Nonunitary}).

A \emph{simple-current} is a primary $\lambda$ for which there exists a permutation $J$ of $P_+^k(X_r^{(i)})$ such that  \begin{equation}\label{simpcurrdefn} N_{\lambda, \mu}^\nu = \delta_{\nu, J\mu} \end{equation} with $\lambda = J0$.  We will also call the permutation $J$ a simple-current.  The simple-currents are precisely those primaries for which equality occurs in (\ref{unitaryvacuum}); they form an Abelian group, which we denote by $\mathcal{J}$.  For each $J \in \mathcal{J}$, there exists a rational number $Q_J(\mu)$ for each $\mu \in P_+^k(X_r^{(i)})$, such that \cite{SY} \begin{equation}\label{SsymwrtQchargea} S_{J\lambda, \mu} = \exp[2 \pi \ii Q_J(\mu)] S_{\lambda\mu} \ . \end{equation} The number $Q_J(\mu)$ has the expression $Q_J(\mu) = h(\mu) + h(J) - h(J\mu)$ (mod $\mathbb{Z}$) in terms of conformal weights.  In all cases except for $E_8^{(1)}$ at level 2, the simple-current group is isomorphic to the centre of the (universal cover of the) corresponding Lie group, and the simple-currents correspond to automorphisms of the extended Dynkin diagram.\footnote{However, not all symmetries of the extended diagram are simple-currents.}
An expression for the $S$-matrix of a nontwisted affine Kac-Moody algebra $X_r^{(1)}$ \cite{KP} is \begin{equation}\label{Sdetform} S_{\lambda\mu} = \kappa^{-r/2}s \sum_{w\in \overline{W}} (\det w) \exp \left[-2\pi \ii \frac{w(\overline{\lambda + \rho}) \cdot (\overline{\mu + \rho})}{\kappa}\right], \end{equation} where $\overline{W}$ is the $X_r$ Weyl group, $\overline{\lambda} = (\lambda_1, \dots , \lambda_r)$, $\overline{\rho} = (1, \dots ,1)$ is the Weyl vector, and $\kappa$ and $s$ are constants depending on $r$ and $k$.  We are using bars to emphasize that the quantities are of the underlying finite dimensional simple algebra $X_r$.  The $S$-matrix is closely related to the $X_r$ characters $\mbox{ch}_{\overline{\lambda}}$ evaluated at elements of finite order, via the Weyl character formula (see e.g. Chapter 10 of \cite{Kac}), and this is the key to our fixed point factorisation formulas.  More precisely, \begin{equation}\label{Sratio} \chi_\lambda(\mu) := \frac{S_{\lambda\mu}}{S_{0\mu}} = \mbox{ch}_{\overline{\lambda}} \left(-2\pi \ii \ \frac{(\overline{\mu + \rho})}{\kappa}\right)  \ . \end{equation}

The characters of a Lie algebra form a ring, called the \emph{character ring}.  A classical result (see e.g. Chapitre IV-VI of \cite{Bour}) is that the character ring of $X_r$ is generated by the characters at the fundamental weights $\overline{\Lambda}_n$; that is, for any integrable highest weight representation $\overline{\lambda}$ of $X_r$, $\mbox{ch}_{\overline{\lambda}}$ is some polynomial $P_{\overline{\lambda}}$ in the variables $\mbox{ch}_{\overline{\Lambda}_n}$.  Due to \eqref{Sratio}, this reduces the question of the existence of a fixed point factorisation at $S_{\lambda \vphi}$ to that of one at $S_{\Lambda_n, \vphi}$, where $\Lambda_n = (k-1) \overline{\Lambda}_0 + \overline{\Lambda}_n$, and so this is our focus.

\subsection{Specific data for the algebras}\label{sApp}

In this section, we give specialised data for the classical WZW models, as well as relevant data for the twisted algebras.\footnote{We use the notation of Kac \cite{Kac} for the twisted algebras.}  Much of the information is presented in the form of tables, for ease of presentation.  We will use $\lambda$ to denote both $X_r$ and $X_r^{(1)}$ weights, as it will be clear from the context which is intended.

In many cases, it will be more convenient to use the \emph{orthogonal coordinates} $\lambda[i]$, rather than Dynkin labels, of $\lambda$ --- for the $B$, $D$ ($A$), algebras, these are coordinates of $\lambda$ with respect to an orthonormal basis $\{e_i\}$ of $\mathbb{R}^r$ ($\mathbb{R}^{r+1}$), and for the $C$-series, it is more convenient to use an orthogonal basis $\{e_i\}$ all of whose elements have length $\sqrt{2}$.  We will mostly work with the \emph{shifted orthogonal coordinates}  $(\lambda + \rho)[i] =: \lambda^+[i]$.  They are given for each algebra in Tables \ref{tab:HW} and \ref{tab:HWtw} below, along with the expression for the level $k$ in terms of Dynkin labels, as well as the number $\kappa$ appearing in \eqref{Sdetform}.  In the first row of Table \ref{tab:HW}, $1 \leq i \leq r+1$; in all other rows, $1 \leq i \leq r$, and in Table \ref{tab:HWtw}, $1 \leq i, j \leq r$.

The simple-current groups and their generators are given in Table \ref{tab:SimpCur} below.  The permutation $J_v$ is also a simple-current for $D_r^{(1)}$ when $r$ is odd --- in that case, $J_v=J_s^2$.  When $r$ is even, we also have the simple-current $J_c := J_v + J_s$.  For the $D$-series, we will use the following conjugation (a conjugation is a graph symmetry fixing the zeroth node): \begin{equation}\label{DChargeConj} C_1: (\lambda_0;  \dots ,  \lambda_r) \mapsto (\lambda_0;  \dots , \lambda_{r-2}, \lambda_r, \lambda_{r-1}) \ .  \end{equation} When $r$ is odd, this is the charge-conjugation $C$ in \eqref{moddatsymmetries2} (charge-conjugation for $r$ even is trivial).

\vspace{.5cm}

\begin{table}[h]\begin{center}
\begin{tabular}{llll}
\\[.08cm]
\hline
$\mathfrak{g}$ &  Level $k$  & $\lambda^+[i]$ & $\kappa$ \\
\hline
\\[.01cm]
$A_r^{(1)}$    & $\sum_{\ell=0}^r\lambda_\ell$                                   & $r + 1 - i + \sum_{\ell=i}^r \lambda_\ell$ & $k + r + 1$ \\[.25cm]
$B_r^{(1)}$    & $\lambda_0 + \lambda_1  + 2\sum_{\ell=2}^{r-1} \lambda_\ell + \lambda_r $ & $r + \frac{1}{2} - i + \sum_{\ell=i}^{r-1} \lambda_\ell + \frac{\lambda_r}{2}$  & $k + 2r - 1$  \\[.25cm]
$C_r^{(1)}$    & $\sum_{\ell=0}^r\lambda_\ell$ & $r + 1 - i + \sum_{\ell=i}^r \lambda_\ell$ & $k+r+1$ \\[.25cm]
$D_r^{(1)}$    & $\lambda_0 + \lambda_1 + 2\sum_{\ell=2}^{r-2}\lambda_\ell + \lambda_{r-1} + \lambda_r$ & $r-i + \sum_{\ell = i}^{r-1} \lambda_\ell + \frac{\lambda_r - \lambda_{r-1}}{2}$ & $k + 2r -2$ \\[.25cm]
$A_{2r}^{(2)}$ & $\lambda_0 + 2\sum_{\ell=1}^r\lambda_\ell$ & $r + 1 - i + \sum_{\ell = i}^r \lambda_\ell$ & $k + 2r + 1$ \\[.25cm]
\hline
\cline{1-4}
\end{tabular}\end{center}\caption{\label{tab:HW} Affine algebra data --- symmetric $S$-matrix}
\end{table}

\vspace{.5cm}

\begin{table}[h]\begin{center}
\begin{tabular}{llll}
\\[.08cm]
\hline
$\mathfrak{g}$ &  Level $k$  & $\nu^+[i]$, $\lambda^+[j]$ & $\kappa$ \\
\hline
\\[.01cm]
$A_{2r-1}^{(2)}$ & $\nu_0 + \nu_1 + 2\sum_{\ell=2}^r\nu_\ell$                 & $r + 1 - i + \sum_{\ell = i}^r \nu_\ell$ & $k + 2r$               \\[.25cm]
                 & $\lambda_0 + 2\sum_{\ell=1}^{r-1}\lambda_\ell + \lambda_r$ & $2r + 1 - 2j + 2\sum_{\ell = j}^{r-1} \lambda_\ell + \lambda_r$ & \\[.25cm]
$D_{r+1}^{(2)}$  & $\nu_0 + 2\sum_{\ell=1}^{r-1}\nu_\ell + \nu_r$             & $2r+1 - 2i + 2\sum_{\ell=i}^{r-1} \nu_\ell + \nu_r$ & $k+2r$      \\[.25cm]
                 & $\lambda_0 + \lambda_1 + 2\sum_{\ell=2}^r\lambda_\ell$     & $r+1-j + \sum_{\ell=j}^r \lambda_\ell$ &                          \\[.25cm]
\hline
\cline{1-4}
\end{tabular}\end{center}\caption{\label{tab:HWtw} Affine algebra data --- nonsymmetric $S$-matrix}
\end{table}

\vspace{.5cm}

\begin{table}[h]\begin{center}
\begin{tabular}{lll}
\\[.08cm]
\hline
$\mathfrak{g}$ & $\mathcal{J}$ & Generators of $\mathcal{J}$ \\
\hline
\\[.01cm]
$A_r^{(1)}$            & $\mathbb{Z}_{r+1}$                  & $J: \lambda \mapsto (\lambda_r; \lambda_0, \dots ,\lambda_{r-1})$           \\[.25cm]
$B_r^{(1)}$            & $\mathbb{Z}_2$                      &  $J: \lambda \mapsto (\lambda_1; \lambda_0, \lambda_2, \dots ,\lambda_r)$   \\[.25cm]
$C_r^{(1)}$            & $\mathbb{Z}_2$                      & $J: \lambda \mapsto (\lambda_r;  \dots  , \lambda_0)$    \\[.25cm]
$D_r^{(1)}$, $r$ odd   & $\mathbb{Z}_4$                      & $J_s: \lambda \mapsto (\lambda_{r-1}; \lambda_r, \lambda_{r-2}, \dots,  \lambda_0)$\\[.25cm]
$D_r^{(1)}$, $r$ even  & $\mathbb{Z}_2 \times \mathbb{Z}_2$  & $J_s: \lambda \mapsto (\lambda_r; \dots ,  \lambda_0)$                       \\[.25cm]
                       &        & $J_v: \lambda \mapsto (\lambda_1; \lambda_0, \lambda_2,  \dots , \lambda_{r-2}, \lambda_r, \lambda_{r-1})$\\[.25cm]
$A_{2r-1}^{(2)}$       &   $\mathbb{Z}_2$                       & $J: \nu \mapsto (\nu_1; \nu_0, \nu_2, \dots , \nu_r)$     \\[.25cm]
$D_{r+1}^{(2)}$        &   $\mathbb{Z}_2$                       &   $J: \nu \mapsto  (\nu_r; \dots , \nu_0)$      \\[.25cm]
\hline
\cline{1-3}
\end{tabular}\end{center}\caption{\label{tab:SimpCur} Simple-current groups and their generators}
\end{table}

The characters of the fundamental weights of $X_r$ can be expressed in terms of elementary symmetric polynomials (see e.g. Chapter 23 of \cite{FH}).  The $d\nth$ elementary symmetric polynomial $E_d$ in variables $a_1, \dots , a_n$, is defined as \begin{equation}\label{Esym} E_d(a_1, \dots , a_n) = \sum_{1 \leq i_1 < \cdots < i_d \leq n} a_{i_1} \cdots a_{i_d} \ , \end{equation} where $E_0(a_1, \dots , a_n) = 1$, $E_{-d}(a_1, \dots , a_n) = 0$ if $d>0$, and $E_d(a_1, \dots , a_n) = 0$ if $n < d$.  We also define $E_d(\emptyset)$ to be the polynomial in zero variables, with $E_d(\emptyset) = 0$ if $d \neq 0$ and $E_0(\emptyset) = 1$.

Explicit expressions for the $S$-matrix \eqref{Sdetform} for the nontwisted algebras can be found in \cite{Alg}; an expression for the $A_{2r}^{(2)}$ $S$-matrix is due to \cite{FSS}.  In the case of $A_{2r-1}^{(2)}$ and $D_{r+1}^{(2)}$, the $S$-matrix is not symmetric --- explicit expressions for these appear in \cite{GG}.  We now remark on the latter two algebras.  The rows and columns of $S$ are indexed by $(r+1)$-tuples $\nu$ and $\lambda$ respectively (see Table \ref{tab:HWtw}).  We set \begin{equation}\label{Sratiotw} \chi_{\lambda}(\nu) := S_{\nu \lambda}/S_{\nu 0} \ , \end{equation} where $\nu$, $\lambda$ are as in Table \ref{tab:HWtw}.  Simple-currents act on the weights $\nu$.

We are interested in the ratios \eqref{Sratio}, \eqref{Sratiotw} of the $S$-matrix.  Evaluating these yields the following expressions in terms of elementary symmetric polynomials:

\vspace{.5cm}
\noindent \emph{The algebra $B_r^{(1)}$.} \begin{equation}\label{Bcharsym} \chi_{\Lambda_n}(\mu) = \left\{ \begin{array}{ll} E_n(1, z_1, \dots , z_r, z_1^{-1}, \dots , z_r^{-1}) & \mbox{ if } 1 \leq n \leq r-1 \\ E_r( z_1^{1/2} + z_1^{-1/2}, \dots , z_r^{1/2} + z_r^{-1/2}) & \mbox{ if } n=r \end{array} \right. \ , \end{equation} where $z_j = \exp\left[-2\pi \ii \frac{\mu^+[j]}{\kappa}\right]$.

\vspace{.5cm}
\noindent\emph{The algebra $C_r^{(1)}$.} \begin{equation}\label{Ccharsym} \chi_{\Lambda_n}(\mu) = E_n(z_1, \dots ,z_r, z_1^{-1}, \dots ,z_r^{-1}) - E_{n-2}(z_1, \dots ,z_r, z_1^{-1}, \dots ,z_r^{-1}) \ , \end{equation} where $z_j = \exp\left[- \pi \ii\frac{\mu^+[j]}{\kappa}\right]$.

\vspace{.5cm}
\noindent\emph{The algebra $D_r^{(1)}$.} \begin{equation}\label{Dsymchar} \chi_{\Lambda_n}(\mu) = \left\{ \begin{array}{ll} E_n(z_1, \dots , z_r, z_1, \dots , z_r^{-1}) & \mbox{ if } 0 \leq n \leq r-2 \\ \sum z_1^{\pm 1/2} \cdots z_r^{\pm 1/2} \mbox{ \small{for an odd number of minus signs}} &  \mbox{ if } n = r-1 \\ \sum z_1^{\pm 1/2} \cdots z_r^{\pm 1/2} \mbox{ \small{for an even number of minus signs}} & \mbox { if } n = r  \end{array} \right. \ , \end{equation} where $z_j = \exp\left[-2 \pi \ii \frac{\mu^+[j]}{\kappa}\right]$.

\vspace{.5cm}
\noindent\emph{The algebra $A_{2r}^{(2)}$.} \begin{equation} \label{AtwSymChar} \chi_{\lambda}(\mu) = E_n(z_1, \dots , z_r, z_1^{-1}, \dots ,z_r^{-1}) - E_{n-2}(z_1, \dots , z_r, z_1^{-1}, \dots , z_r^{-1}) \ , \end{equation} where $z_j := \exp\left[-2 \pi \ii \frac{\mu^+[j]}{\kappa}\right]$.

\vspace{.5cm}
\noindent\emph{The algebra $A_{2r-1}^{(2)}$.} \begin{equation}\label{Atw2charsym} \chi_{\Lambda_n}(\nu) = \left\{ \begin{array}{ll} E_n(1, z_1, \dots , z_r, z_1^{-1}, \dots , z_r^{-1}) & \mbox{ if } 0 \leq n \leq r-1 \\ E_r( z_1^{1/2} + z_1^{-1/2}, \dots , z_r^{1/2} + z_r^{-1/2}) & \mbox{ if } n=r \end{array} \right. \ , \end{equation} where $z_i = \exp\left[-2\pi \ii \frac{\nu^+[i]}{\kappa}\right]$.

\vspace{.5cm}
\noindent\emph{The algebra $D_{r+1}^{(2)}$.} \begin{equation}\label{Dtwcharsym} \chi_{\Lambda_n}(\nu) = E_n(z_1, \dots , z_r, z_1^{-1}, \dots , z_r^{-1}) - E_{n-2}(z_1, \dots , z_r, z_1^{-1}, \dots , z_r^{-1}) \ , \end{equation} where $z_i = \exp\left[-\pi \ii \frac{\nu^+[i]}{\kappa}\right]$.


\section{The fixed point factorisation formulas}\label{sfpf}
\setcounter{equation}{0}

We will prove the following theorem:

\begin{theorem}\label{thm:FPF} Let $S$ be the modular $S$-matrix for the WZW model corresponding to $X_{r}^{(i)}$, level $k$, where $X_r^{(i)}$ is given in the first column of Table \ref{table:fpf}.  Let $\mathcal{J}$ be its group of simple-currents, and let $\vphi$ be a primary for $X_r^{(i)}$ that is fixed by a subgroup of $\mathcal{J}$.  Then for any $\lambda \in P_+^k(X_r^{(i)})$, $S_{\vphi \lambda}$ can be expressed in terms of the $S$-matrix for the algebra $Y_{s}^{(j)}$ at level $\ell$, given in the third and fourth columns of Table \ref{table:fpf}.  Explicit formulas for the case that $\lambda$ is a fundamental weight are given in \eqref{FPFAfundweights}, \eqref{BFPF}, \eqref{Coddfundweightfpf}, \eqref{Crevfpf}, \eqref{FPFDJv}, \eqref{Droddfpfeqn}, \eqref{Drevenfpfeqn}, \eqref{Drevenfpfvseqn}, \eqref{Atwfpfeqn}, \eqref{Dtwfpfeqn} and \eqref{fpfeqnDtwev}.
\end{theorem}

\begin{remark} The proof of eqn \eqref{FPFAfundweights} was done in \cite{FPF}. \end{remark}

The proof of Theorem \ref{thm:FPF} is surprisingly straightforward, once the challenge of realising that a fixed point factorisation exists and what it will look like has been overcome.  Earlier work on this was done in \cite{ThesisP}.  The key to the proof is the relationship \eqref{Sratio} between the $S$-matrix and the characters of the underlying simple finite dimensional algebra, and the fact that the latter (at the fundamental weights) can be expressed through symmetric polynomials.

The table below summarizes the algebras, their simple-current group generators, and the smaller-rank algebras, which we call the `fixed point factorisation (FPF) algebra', involved in each case.
\vspace{.5cm}
\begin{table}[h]\begin{center}
\begin{tabular}{lcll}
\\[.08cm]
\hline
$X_r^{(i)}$, level $k$ & Simple-current & FPF algebra & Level   \\
\hline
\\[.01cm]
$A_r^{(1)}$       &   $J^d$   & $A_{d-1}^{(1)}$ & $\frac{kd}{r+1}$ \\[.25cm]
$B_r^{(1)}$       &   $J$   &   $A_{2(r-1)}^{(2)}$  & $k$  \\[.25cm]
$C_r^{(1)}$, $r$ odd    &   $J$     &   $C_{\frac{r-1}{2}}^{(1)}$   & $\frac{k}{2}$ \\[.25cm]
$C_r^{(1)}$, $r$ even   &   $J$     &   $A_{2\left(\frac{r}{2}\right)}^{(2)}$   & $k$ \\[.25cm]
$D_r^{(1)}$       &   $J_v$   &   $C_{r-2}^{(1)}$   & $\frac{k}{2}$ \\ [.25cm]
$D_r^{(1)}$, $r$ odd   &   $J_s$   &   $C_{\frac{r-3}{2}}^{(1)}$   & $\frac{k}{4}$ \\[.25cm]
$D_r^{(1)}$, $r$ even  &   $J_s$   &   $B_{\frac{r}{2}}^{(1)}$   &  $\frac{k}{2}$ \\[.25cm]
$A_{2r-1}^{(2)}$  &   $J$   &   $C_{r-1}^{(1)}$   &  $\frac{k}{2}$ \\[.25cm]
$D_{r+1}^{(2)}$, $r$ odd  &   $J$   &   $A_{2\left(\frac{r-1}{2}\right)}^{(2)}$   &  $\frac{k}{2}$ \\[.25cm]
$D_{r+1}^{(2)}$, $r$ even & $J$ & $D_{r+1}^{(2)}$ & $\frac{k}{2}$ \\[.25cm]
\hline
\cline{1-4}
\end{tabular}\end{center}\caption{\label{table:fpf} Fixed point factorisation algebras for the classical affine algebras}
\end{table}

\subsection{Useful facts}\label{subsec:UsefulFacts}

We collect here some basic results which we will use in the proofs of the fixed point factorisation formulas, before turning to the formulas themselves.  The first is a convenient source of zeros in the $S$-matrix, and the remaining two are properties of elementary symmetric polynomials.

Let $\vphi$ be a fixed point for a simple-current $J$.  Then (\ref{SsymwrtQchargea}) yields the symmetry $ S_{\vphi \lambda} = \exp[2\pi \ii Q_J(\lambda)] S_{\vphi \lambda}$, which in turn yields the useful fact: \begin{fact}\label{fact:charge} $S_{\vphi \mu} = 0$ whenever $Q_J(\mu) \not \in \mathbb{Z}$.\end{fact}

Recall the definition of $E_d$ in \eqref{Esym}.  The following two equations are immediate consequences of the definition: \begin{fact}\label{Erecursion} For any variables $a, b_1, \dots , b_n$, \begin{eqnarray} E_d(a, b_1, \dots , b_n) &=& aE_{d-1}(b_1, \dots ,b_n) + E_d(b_1 , \dots b_n) \ , \nonumber \\ E_d(a, -a, b_1, \dots b_n ) &=& -a^2E_{d-2}(b_1, \dots , b_n) + E_d(b_1, \dots b_n) \ .  \nonumber \end{eqnarray} \end{fact}

\begin{fact}\label{lemma:Eeven} Let $m \geq 0$.  Then \begin{equation}\nonumber E_{2m}(a_1, -a_1, a_1^{-1}, -a_1^{-1}, \dots ,a_n, -a_n, a_n^{-1}, -a_n^{-1}) = (-1)^m E_m(a_1^2, a_1^{-2}, \dots a_n^2, a_n^{-2}) \ , \end{equation} for any variables $a_1, \dots ,a_n$.
\end{fact}

\noindent\emph{Proof of Fact \ref{lemma:Eeven}.} Let $d=2m$, and define $E_{(j; \ell)} := E_j(a_\ell, -a_\ell, a_\ell^{-1}, -a_\ell^{-1})$, and $E'_{(j; \ell)} := E_j(a_\ell^2, a_\ell^{-2})$.  Then $E_{(0; \ell)} = E_{(4; \ell)} = 1$; $E_{(1; \ell)} = E_{(3; \ell)} = 0$; $E_{(2; \ell)} = -E'_{(1; \ell)}$, and $E_{(j; \ell)}$ is zero for $j > 4$.  Thus, for each $j \in \{0, 2, 4\}$, $E_{(j; \ell)} = (-1)^{j/2} E'_{(\frac{j}{2}; \ell)}$ .

Let $[d_1, \dots , d_n]$ be an ordered partition of $d$.  We will be interested only in those partitions with $d_i \in \{0, 2, 4\}$.  Let $d':=d/2$ and $d'_i = d_i/2$.  Let $E_d$ denote the $d\nth$ elementary symmetric polynomial in variables $a_1$, $-a_1$, $a_1^{-1}$, $-a_1^{-1}, \dots, a_n$, $-a_n$, $a_n^{-1}$, $-a_n^{-1}$.  Then the calculation \begin{eqnarray} E_d &=& \sum_{\stackrel{[d_1, \dots , d_n]}{d_i \in \{0,2,4\}}} E_{(d_1; 1)} \cdots E_{(d_n; n)} \nonumber \\ &=& (-1)^{d'} \sum_{\stackrel{[d'_1, \dots , d'_n]}{d'_i \in \{0,1,2\}}} E'_{(d'_1; 1)} \cdots E'_{(d'_n; n)} \nonumber \\ &=& E_{d'}(a_1^2, a_1^{-2}, \dots , a_n^2, a_n^{-2})  \nonumber \end{eqnarray} establishes the result. $\square$


\subsection{The algebra $A_r^{(1)}$}\label{ssA}

For completeness, we summarize the $A_r^{(1)}$ fixed point factorisation.  This was done in \cite{FPF}, where a fixed point factorisation was found for $S_{\lambda \vphi}$ for all weights $\lambda$.  We include the formula for the case that $\lambda$ is a fundamental weight.

Let $n:= r+1$.  The Coxeter-Dynkin diagram for $A_r^{(1)}$ is a regular $n$-gon.  Let $J$ be the rotation of $2\pi/n$ radians: the group of simple-currents is the group $\langle J \rangle \cong \mathbb{Z}_{n}$ generated by $J$.  There is a fixed point factorisation for $J^d$ where $d|n$, in which case the order of $J^d$ is $n/d$.  The case $d=n$ is trivial ($J = id$), and the case $d=1$ is degenerate (the fixed point factorisation formulas work with $S' = \chi' = 1$).  The $A$-series is the only case where the order of the simple-current group grows with the rank of the algebra.

Suppose $d|n$, and let $\vphi \in P_+^k(A_r^{(1)})$ be a fixed point for $J^d$.  Define the truncated weight \begin{equation}\label{Aphiprime} \widetilde{\vphi} = (\vphi_0; \vphi_1, \dots , \vphi_{d - 1}) \ \end{equation} it lies in $P_+^{\frac{kd}{n}}(A_{d - 1}^{(1)})$, and the fixed point factorisation involves $\frac{n}{d}$ copies of the $S$-matrix for $A_{d-1}^{(1)}$ at level $kd/n$.  The \emph{$n$-ality} of a weight $\lambda$ is defined by $t(\lambda) = \sum_{i=1}^r i\lambda_i$.  By Fact \ref{fact:charge}, $S_{\lambda\vphi} = 0$ whenever $t(\lambda) \not \equiv 0$ (mod $\frac{n}{d}$).  We then have the following: \begin{fpf} Let $n=r+1$, and let $d$ be a proper divisor of $n$.  Let $\vphi$ be fixed by the subgroup $\langle J^{n/d} \rangle$ of the simple-current group, and let $\wtild{\vphi}$ be as in \eqref{Aphiprime}.  We then have \begin{equation}\label{FPFAfundweights} \chi_{\Lambda_\ell}(\vphi) = \left\{ \begin{array}{ll} \chi'_{\Lambda'_{\frac{\ell d}{n}}}(\wtild{\vphi}) & \mbox{if } \frac{n}{d}  | \ell \\  0 & \mbox{if } \frac{n}{d} \nmid \ell \end{array}\right. \ , \end{equation} where primes denote $A_{d-1}^{(1)}$ level $kd/n$ quantities. \end{fpf}


\subsection{The algebra $B_r^{(1)}$}\label{ssB}

A fixed point is of the form \begin{equation}\label{Bfixedpointphi} \vphi = (\vphi_1; \vphi_1, \vphi_2, \dots ,\vphi_r) \ ; \end{equation} they satisfy \begin{equation}\label{Bweightaddupfixed} 2(\vphi_1 + \cdots + \vphi_{r-1}) + \vphi_r = k \ . \end{equation}

Let \begin{equation}\label{Atwistevenphiprime} \wtild{\vphi} = (\vphi_r; \vphi_{r-1}, \dots ,\vphi_1) \ . \end{equation}  Then by \eqref{Bweightaddupfixed}, $\wtild{\vphi} \in P_+^k\left(A_{2(r-1)}^{(2)}\right)$.

\begin{fpf}\label{prop:BFPF} Let $0 \leq n \leq r$, and let $\vphi$, $\wtild{\vphi}$ be as in (\ref{Bfixedpointphi}), (\ref{Atwistevenphiprime}) respectively.  Then \begin{equation}\label{BFPF} \chi_{\Lambda_n}(\vphi) = \left\{ \begin{array}{ll} (-1)^n \left(\chi'_{\Lambda'_n}(\wtild{\vphi}) + \chi'_{\Lambda'_{n-1}}(\wtild{\vphi}) \right) & \mbox{if } 0 \leq n < r \\ 0 & \mbox{if } n = r. \end{array} \right. , \end{equation} where primes denote $A_{2(r-1)}^{(2)}$ level $k$ quantities.
\end{fpf}

\noindent\emph{Proof.} By Fact \ref{fact:charge}, $\chi_{\Lambda_r}(\vphi) = 0$.  Let $n \in \{0, \dots ,r-1\}$, and define $\vphi^j := j + \sum_{\ell=1}^j \varphi_\ell$ , with $\vphi^0 = 0$.  By Equation \eqref{Bweightaddupfixed}, the shifted orthogonal coordinates of $\vphi$ are given by $\vphi^+[j] = \kappa/2 - \vphi^{j-1}$, for $1 \leq j \leq r$.  Let $\xi_j := \exp\left[2 \pi \ii \frac{\vphi^j}{\kappa}\right]$, for $1 \leq j \leq r-1$.  Then putting $\mu = \vphi$ into \eqref{Bcharsym}, we have $z_1 = -1$, and $z_j = -\xi_{j-1}$, for $2 \leq j \leq r$.  Thus, $\chi_{\Lambda_n}(\vphi)$ is the $n\nth$ elementary symmetric polynomial in variables $1, -1, -1$, $-\xi_1^{\pm 1}, \dots ,-\xi_{r-1}^{\pm 1}$.   Let $\veps_n := E_n(\xi_1, \dots , \xi_{r-1}, \xi_1^{-1}, \dots , \xi_{r-1}^{-1})$.  Then applying Fact \ref{Erecursion} (and factoring out $(-1)^n$) to $\chi_{\Lambda_n}(\vphi)$ yields: \begin{equation}\label{Bcharclean} \chi_{\Lambda_n}(\vphi) = (-1)^n (\veps_n + \veps_{n-1} - \veps_{n-2} - \veps_{n-3}) \ . \end{equation}

The shifted orthogonal coordinates of $\wtild{\vphi}$ are $\vphi^+[r-j] = \vphi^j$, for $1 \leq j \leq r-1$.  By \eqref{AtwSymChar}, the $A_{2(r-1)}^{(2)}$ level $k$ $S$-ratios are given by $\chi'_{\Lambda'_n}(\wtild{\vphi}) = \veps_n - \veps_{n-2}$. $\square$


\subsection{The algebra $C_r^{(1)}$}\label{ssC}

The fixed points and fixed point factorisation algebra depend on whether $r$ is odd or even, and are given in each case.  Throughout this section, we let $\vphi^j = j + \sum_{\ell=0}^{j-1} \vphi_\ell$ , $\zeta_j = \exp\left[\pi \ii \frac{\vphi^+[j]}{\kappa}\right]$ and $\xi_j = \zeta_j^2$.

\paragraph{When $r$ is odd.}

Fixed points are of the form \begin{equation}\label{Coddphi} \vphi = (\vphi_0; \vphi_1, \dots, \vphi_{(r-1)/{2}}, \vphi_{(r-1)/{2}}, \dots, \varphi_1, \vphi_0) \ , \end{equation} with \begin{equation}\label{Coddfixedweightaddup} \vphi_0 + \cdots + \varphi_{(r-1)/{2}} = k/2 \ . \end{equation}  Let \begin{equation}\label{Coddphiprime} \wtild{\vphi} = (\vphi_{(r-1)/{2}}; \dots , \varphi_1, \vphi_0) \ . \end{equation}  By \eqref{Coddfixedweightaddup}, $\wtild{\vphi} \in P_+^{\frac{k}{2}}\left(C_{\frac{r-1}{2}}^{(1)}\right)$.

\begin{fpf}\label{corollary:CoddFundFPF} Let $r$ be odd, and let $m \in \{ 0, \dots , \frac{r-1}{2}\}$.  Then \begin{equation}\label{Coddfundweightfpf}
\chi_{\Lambda_{n}}(\vphi) = \left\{ \begin{array}{ll} (-1)^m \chi'_{\Lambda'_m}(\wtild{\vphi}) & \mbox{ if } n=2m \\ 0 & \mbox{ if } n = 2m+1 \end{array} \right. \ , \end{equation} where $\wtild{\vphi}$ is as in \eqref{Coddphiprime}, and primes denote $C_{\frac{r-1}{2}}^{(1)}$ level $k/2$ quantities. \end{fpf}

\noindent\emph{Proof.}  Fact \ref{fact:charge} implies $\chi_{\Lambda_n}(\vphi) = 0$ whenever $n$ is odd. Let $r' = (r-1)/2$.  By Equation \eqref{Coddfixedweightaddup}, the shifted orthogonal coordinates of $\vphi$ are \begin{equation}\label{Coddorthcoords} \vphi^+[j] = \kappa - \vphi^j \ \ \ \ ; \ \ \ \  \vphi^+[r+1-j] = \vphi^j \end{equation} for $j = 1, \dots , r'$ , and $\vphi^+[(r+1)/2] = \kappa/2$.  Substituting $\mu = \vphi$ in Equation \eqref{Ccharsym} yields $z_j = -\zeta_j$, $z_{r+1-j} = \zeta_j^{-1}$ for $1 \leq j \leq r'$, and $z_{\frac{r+1}{2}} = -\ii$.

Let $E_n$ be the $n\nth$ elementary symmetric polynomial in the variables $\ii$, $-\ii$, $\pm \zeta_j$, $\pm \zeta_j^{-1}$, for $j \in \{1, \dots , r'\}$, and $\veps_n := E_n(\xi_1, \xi_1^{-1}, \dots , \xi_{r'}, \xi_{r'}^{-1})$.  Then Facts \ref{Erecursion} and \ref{lemma:Eeven} yield \begin{equation}\label{Coddcharexp} \chi_{\Lambda_{2m}}(\vphi) = E_{2m} - E_{2(m-1)} = (-1)^m (\veps_m - \veps_{m-2}) \ . \end{equation}

Consider $C_{r'}^{(1)}$ at level $k/2$.  The shifted orthogonal coordinates of $\wtild{\vphi}$ are $\wtild{\vphi}^+[r'+1 - j] = \vphi^j$, for $1 \leq j \leq r'$.  The $m\nth$ elementary symmetric polynomial in variables $z'_j = \exp\left[\pi \ii \frac{\wtild{\vphi}^+[j]}{\kappa'}\right]$ (where $\kappa' = \kappa/2$), is $\veps_m$.  Thus, by \eqref{Ccharsym}, \begin{equation}\label{Coddprimcharexp} \chi'_{\Lambda'_m}(\wtild{\vphi}) = \veps_m - \veps_{m-2}  \ . \end{equation}  The fixed point factorisation formula \eqref{Coddfundweightfpf} follows from \eqref{Coddcharexp} and \eqref{Coddprimcharexp}. $\square$

\paragraph{When $r$ is even.}

When $r$ is even, there is a fixed point factorisation at all levels $k$.  The fixed points of $J$ are weights \begin{equation}\label{evenphi}\vphi = (\vphi_0;  \dots, \vphi_{\frac{r}{2} - 1}, \vphi_{\frac{r}{2}}, \vphi_{\frac{r}{2}-1}, \dots ,  \vphi_0) \ ,\end{equation} where  \begin{equation}\label{Cevenphiweightaddup} 2(\vphi_0 + \dots + \vphi_{\frac{r}{2} - 1}) + \vphi_{\frac{r}{2}} = k \ .\end{equation}

Let \begin{equation}\label{Cevphiprime} \wtild{\vphi} = (\vphi_{\frac{r}{2}}; \dots, \vphi_0) \ . \end{equation} By \eqref{Cevenphiweightaddup}, $\wtild{\vphi} \in P_+^k(A_{2\left(\frac{r}{2}\right)}^{(2)})$.  We have: \begin{fpf}\label{theorem:Crevenfpf}
Let $r$ be even, and let $\vphi$, $\wtild{\vphi}$ be as in \eqref{evenphi}, \eqref{Cevphiprime} resp.  Let $0 \leq m \leq \frac{r}{2}$.  Then \begin{equation}\label{Crevfpf} \chi_{\Lambda_{2m}}(\vphi) = (-1)^m \sum_{\ell=0}^m \chi'_{\Lambda'_\ell}(\wtild{\vphi}) \ , \end{equation} where primes denote $A_{2\left(\frac{r}{2}\right)}^{(2)}$ level $k$ quantities. \end{fpf}

\noindent\emph{Proof.} The shifted orthogonal coordinates of $\vphi$ are given in \eqref{Coddorthcoords}, for $1 \leq j \leq \frac{r}{2}$.  Putting $\mu = \varphi$ into \eqref{Ccharsym}, the set of variables $\{z_j, z_j^{-1} \ | \ 1 \leq j \leq r \}$, is the set $\{\zeta_j, \zeta_j^{-1}, -\zeta_j, -\zeta_j^{-1} \ | \ 1 \leq j \leq \frac{r}{2} \}$.  Equation \eqref{Ccharsym} and Fact \ref{lemma:Eeven} then imply that \begin{equation}\label{Cevcharclean} \chi_{\Lambda_{2m}}(\vphi) = (-1)^m(\veps_m + \veps_{m-1}) \ , \end{equation} where $\veps_m := E_m(\xi_1, \dots , \xi_{\frac{r}{2}}, \xi_1^{-1}, \dots , \xi_{\frac{r}{2}}^{-1})$.

Let $r' = r/2$.  Consider the algebra $A_{2\left(\frac{r}{2}\right)}^{(2)}$ level $k$.  The shifted orthogonal coordinates of $\wtild{\vphi}$ are $\wtild{\vphi}^+[r' + 1 - j] = \vphi^j$, for $1 \leq j \leq r'$.  Let $z'_j := \exp\left[-2 \pi \ii \frac{\wtild{\vphi}^+[j]}{\kappa}\right]$ (here, $\kappa' = \kappa$).  Then the set $\{z'_j, {z'}_j^{-1}\}_{1 \leq j \leq r/2} = \{\xi_j, \xi_j^{-1}\}_{1 \leq j \leq r/2}$ , so by \eqref{AtwSymChar}, \begin{equation}\label{Cevcharprimclean} \chi'_{\Lambda'_m}(\wtild{\vphi}) = \eps_m - \eps_{m-2} \ . \end{equation}  Equations \eqref{Cevcharclean} and \eqref{Cevcharprimclean} now establish \eqref{Crevfpf}. $\square$


\subsection{The algebra $D_r^{(1)}$}\label{ssD}

The nontrivial simple-currents are $J_v$, $J_s$ and $J_vJ_s$ (see Table \ref{tab:SimpCur}).  Throughout this section, we let $\xi_j = \exp\left[2\pi \ii \frac{\varphi^j}{\kappa}\right]$, with the relevant $\kappa$ and $\varphi^j$ (given in each case).

\paragraph{The simple-current $J_v$.}
The simple-current $J_v$ has fixed points \begin{equation}\label{DJvfixedpt} \vphi = (\vphi_1; \vphi_1, \vphi_2, \dots, \varphi_{r-2}, \vphi_{r-1}, \vphi_{r-1}) \ , \end{equation} with \begin{equation}\label{DJvfixedptweightaddup} \vphi_1 + \dots + \varphi_{r-1} = \frac{k}{2} \ .\end{equation}

Define \begin{equation}\label{DJvphiprime} \wtild{\vphi} := (\vphi_{r-1}; \vphi_{r-2}, \dots , \vphi_1) \ .\end{equation}  Then $\wtild{\vphi} \in P_+^{\frac{k}{2}}(C_{r-2}^{(1)})$.  The fixed point factorisation in this case is:

\begin{fpf}\label{proposition:FPFDJv} Let $1 \leq n \leq r$, and let $\vphi$, $\wtild{\vphi}$ be as in \eqref{DJvfixedpt}, \eqref{DJvphiprime} respectively.  Then \begin{equation}\label{FPFDJv} \chi_{\Lambda_n}(\vphi) = \left\{ \begin{array}{ll} (-1)^n \left( \chi'_{\Lambda'_n} (\wtild{\vphi}) -  \chi'_{\Lambda'_{n-2}} (\wtild{\vphi}) \right) & \mbox{if } 0 \leq n \leq r-2 \\  0  & \mbox{if } n = r-1, \ r \end{array} \right. \ ,
\end{equation} where primes denote $C_{r-2}^{(1)}$ level $\frac{k}{2}$ quantities.\end{fpf}

\noindent\emph{Proof.} By Fact \ref{fact:charge}, $\chi_{\Lambda_{r-1}}(\vphi) = \chi_{\Lambda_r}(\vphi) = 0$.  Let $\vphi^j = j + \sum_{\ell = 1}^j \varphi_\ell$ , with $\vphi^0 = 0$.  The orthogonal coordinates of the fixed point $\vphi$ are $\vphi^+[j] = \frac{\kappa}{2} - \vphi^{j-1}$, for $1 \leq j \leq r$.  Putting $\mu = \vphi$ into \eqref{Dsymchar} yields $z_1 = -1$, $z_r=1$, and $z_j = -\xi_{j-1}$ for $2 \leq j \leq r-1$.  Thus, $\chi_{\Lambda_n}(\vphi)$ is the $n\nth$ elementary symmetric polynomial in variables $-1$, $-1$, $1$, $1$, and $-\xi_j^{\pm 1}$ for $1 \leq j \leq r-2$.  Let $\veps_n := E_n(\xi_1, \xi_1^{-1}, \dots , \xi_{r-2}, \xi_{r-2}^{-1})$.  Then by Facts \ref{Erecursion} and \ref{lemma:Eeven}, \begin{equation}\label{DJvcharclean}\chi_{\Lambda_n}(\vphi) = (-1)^n (\veps_n - 2\veps_{n-2} + \veps_{n-4}) \ . \end{equation}

Now consider $C_{r-2}^{(1)}$ at level $k/2$.  Here, $\kappa' = \kappa/2$, and the shifted orthogonal coordinates of $\wtild{\vphi}$ are given by $\wtild{\vphi}^+[r-1-j] = \vphi^j$, for $1 \leq j \leq r-2$.  Let ${z'}_j = \exp\left[-\pi \ii \frac{\wtild{\vphi}^+[j]}{\kappa'}\right]$ as in \eqref{Ccharsym}.  Then the set of variables $\{{z'}_1^{\pm 1}, \dots , {z'}_{r-2}^{\pm 1} \}$ is $\{\xi_1^{\pm 1}, \dots , \xi_{r-2}^{\pm 1} \}$, and so $\chi'_{\Lambda'_n}(\wtild{\vphi})$ is the difference $\veps_{n} - \veps_{n-2}$.  This coupled with \eqref{DJvcharclean} establishes \eqref{FPFDJv}. $\square$

\paragraph{The simple-current $J_s$ when $r$ is odd.}

There are fixed points when $4 | k$.  The simple-current $J_s$ has fixed points are \begin{equation}\label{DoddJsfix} \vphi = (\vphi_1; \vphi_1, \vphi_2, \dots, \vphi_{\frac{r-1}{2}}, \vphi_{\frac{r-1}{2}}, \dots , \vphi_2, \vphi_1, \vphi_1) \ , \end{equation} with \begin{equation}\label{DoddJsphiaddup} \vphi_1 + \dots + \vphi_{\frac{r-1}{2}} = \frac{k}{4} \ .\end{equation}  Let \begin{equation}\label{Doddsphiprim} \wtild{\vphi} = ( \vphi_{\frac{r-1}{2}}; \dots , \vphi_1) \ . \end{equation}  Then by \eqref{DoddJsphiaddup}, $\wtild{\vphi} \in P_+^{\frac{k}{4}}\left(C_{\frac{r-3}{2}}^{(1)}\right)$.

Our fixed point factorisation formula is:
\begin{fpf}\label{Droddfpf} Let $r$ be odd, and let $\vphi$, $\wtild{\vphi}$ be as in \eqref{DoddJsfix}, \eqref{Doddsphiprim} respectively.  Then \begin{equation}\label{Droddfpfeqn} \chi_{\Lambda_{n}}(\vphi) = \left\{ \begin{array}{ll} (-1)^m \left( \chi'_{\Lambda'_m}(\wtild{\vphi}) + \chi'_{\Lambda_{m-1}}(\wtild{\vphi}) \right) & \mbox{ if } 0 \leq n=2m \leq r-3 \\ 0 & \mbox{ otherwise} \end{array} \right. \  , \end{equation} where primes denote $C_{\frac{r-3}{2}}^{(1)}$ level $k/4$ quantities. \end{fpf}

\noindent\emph{Proof.}  By Fact \ref{fact:charge}, $\chi_{\Lambda_n}(\vphi) = 0$ whenever $n$ is odd, or $n=r-1$.   Let $\vphi^j := j + \sum_{\ell=1}^{j} \varphi_\ell$.  The shifted orthogonal coordinates of $\vphi$ are:

\begin{equation}\label{Dphiorthcoords}
\vphi^+[j] = \frac{\kappa}{2} - \vphi^{j-1} \ \ \ \ ; \ \ \ \  \vphi^+[r+1-j] = \vphi^{j-1} \ ,
\end{equation} where $1 \leq j \leq \frac{r-1}{2}$, and $\vphi^+[(r+1)/2] = \kappa/4$.

Let $\xi_j := \exp\left[2 \pi \ii \frac{\vphi^j}{\kappa}\right]$.  Then putting $\mu = \vphi$ into \eqref{Dsymchar}, we have $z_1 = -1$, $z_r = 1$, $z_{(r+1)/2} = -\ii$, $z_j = -\xi_{j-1}$, and $z_{r+1-j} = \xi_{j-1}^{-1}$ for $2 \leq j \leq (r-1)/2$.  Let $r':= \frac{r-3}{2}$ .  The $S$-ratio $\chi_{\Lambda_n}(\vphi)$ is the $n\nth$ elementary symmetric polynomial in variables $\pm 1$, $\pm 1$, $\pm \ii$, and $\pm \xi_j$, $\pm \xi_j^{-1}$, for $1 \leq j \leq r'$.  Define $\eps_m = E_m(\xi_1^2, \xi_1^{-2}, \dots , \xi_{r'}^2, \xi_{r'}^{-2})$.  Let $0 \leq m \leq r'$.  Then Facts \ref{Erecursion} and \ref{lemma:Eeven} yield the expression \begin{equation}\label{Dsoddcharclean} \chi_{\Lambda_{2m}}(\vphi) = (-1)^m (\eps_m + \eps_{m-1} - \eps_{m-2} - \eps_{m-3}) \ . \end{equation}

Consider $C_{r'}^{(1)}$ at level $k/4$.  The shifted orthogonal coordinates of $\wtild{\vphi}$ are $\wtild{\vphi}^+[r'+1-j] = \varphi^{j}$, for $1 \leq j \leq r'$.  By \eqref{Ccharsym}, the character $\chi'_{\Lambda'_m}(\wtild{\vphi})$ is the difference $E_m - E_{m-2}$ of symmetric polynomials in the variables $z'_j$ and ${z'_j}^{-1}$, with $z'_j = \exp\left[\pi \ii \frac{\wtild{\vphi}^+[j]}{\kappa'}\right]$, where $\kappa' = \kappa/4$.  That is,  \begin{equation}\label{Dsoddcharprimeclean} \chi'_{\Lambda'_m}(\wtild{\vphi}) = \eps_m - \eps_{m-2} \ . \end{equation}  The result \eqref{Droddfpfeqn} now follows from \eqref{Dsoddcharclean} and \eqref{Dsoddcharprimeclean}. $\square$

\paragraph{The simple-current $J_s$ when $r$ is even.}

The simple-current generator $J_s$ has fixed points when $k$ is even.  They are the set of all weights \begin{equation}\label{DevenJsfixpt} \vphi = (\vphi_0;  \dots, \vphi_{\frac{r}{2} - 1}, \vphi_{\frac{r}{2}}, \varphi_{\frac{r}{2} -1}, \dots , \vphi_0),\end{equation} with \begin{equation}\label{DevenJsphiweightaddup} \vphi_0 + \vphi_1 + 2(\vphi_2 + \dots + \vphi_{\frac{r}{2} -1}) + \vphi_{\frac{r}{2}} = k/2 \ .\end{equation}

Define \begin{equation}\label{DevenJsphiprime} \wtild{\vphi} := (\vphi_0; \vphi_1,  \dots , \vphi_{\frac{r}{2}}) \ . \end{equation} By \eqref{DevenJsphiweightaddup}, $\wtild{\vphi} \in P_+^{\frac{k}{2}}\left(B_{\frac{r}{2}}^{(1)}\right)$.

We have: \begin{fpf} \label{Drevenfpf} Let $r$ be even, and let $\vphi$, $\wtild{\vphi}$ be as in \eqref{DevenJsfixpt}, \eqref{DevenJsphiprime} respectively.  Then \begin{equation}\label{Drevenfpfeqn} \chi_{\Lambda_{n}}(\vphi) = \left\{ \begin{array}{ll} (-1)^{m} \sum_{\ell = 0}^m \chi'_{\Lambda'_\ell}(\wtild{\vphi}) & \mbox{ if } 0 \leq n=2m \leq r-2 \\ (-1)^{\frac{r}{2}}\chi'_{\Lambda'_{\frac{r}{2}}} & \mbox{ if } n = r-1 \mbox{ and } 2 ||r \mbox{ , or } n=r \mbox{ and } 4|r \\ 0 & \mbox{otherwise}  \end{array} \right. \  , \end{equation} where primes denote $B_{\frac{r}{2}}^{(1)}$, level $k/2$ quantities. \end{fpf}

\noindent\emph{Proof.} By Fact \ref{fact:charge}, $\chi_{\Lambda_n}(\vphi) = 0$ whenever $n$ is odd and $1 \leq n \leq r-3$.  If $4|r$, then $\chi_{\Lambda_{r-1}}(\vphi) = 0$, and if $4 \nmid r$, then $\chi_{\Lambda_r}(\vphi) = 0$.  Let $\displaystyle{\vphi^j := j + \sum_{\ell=1}^{j} \vphi_\ell + \frac{\vphi_0 - \vphi_1}{2}}$ .  By \eqref{DevenJsphiweightaddup}, the shifted orthogonal coordinates of $\vphi$ are given by \eqref{Dphiorthcoords}, for $1 \leq j \leq \frac{r}{2}$.  Let $\xi_j = \exp\left[2\pi \ii\frac{\vphi^{j-1}}{\kappa}\right]$.  Putting $\mu = \vphi$ into \eqref{Dsymchar} yields $z_j = -\xi_j$ and $z_{r+1-j} = \xi_j^{-1}$, for $1 \leq j \leq \frac{r}{2}$ .  Let $\eps_m = E_m(\xi_1^2, \xi_1^{-2}, \dots ,\xi_{\frac{r}{2}}^2, \xi_{\frac{r}{2}}^{-2})$, where $m \in \{0, \dots , \frac{r}{2}-1\}$.  Then \eqref{Dsymchar} and Fact \ref{lemma:Eeven} implies \begin{equation}\label{Dsevcharclean} \chi_{\Lambda_{2m}} = (-1)^m\eps_m \ . \end{equation}

Now consider $B_{\frac{r}{2}}^{(1)}$ at level $k/2$.  Here, $\kappa' = \kappa/2$, and the shifted orthogonal coordinates of $\wtild{\vphi}$ are $\wtild{\vphi}^+[j] = \kappa/4 - \vphi^{j-1}$, for $1 \leq j \leq \frac{r}{2}$ .  By \eqref{Bcharsym}, $\chi'_{\Lambda'_m}(\wtild{\vphi})$ is the $m\nth$ elementary polynomial in the variables 1, $-\xi_j^2$, and $-\xi_j^{-2}$, $1 \leq j \leq \frac{r}{2}$ .  Then Fact \ref{Erecursion} yields \begin{equation}\label{Dsevencharprimclean} \chi'_{\Lambda'_m}(\wtild{\vphi}) = (-1)^m(\eps_m - \eps_{m-1}) \ . \end{equation}  Thus for $0 \leq n \leq r-2$, the result \eqref{Drevenfpfeqn} follows from \eqref{Dsevcharclean} and \eqref{Dsevencharprimclean}.

It remains to consider the case $n=r-1, r$.  By \eqref{Dsymchar}, \begin{equation}\label{Dsevlastcharseqn} \chi_{\Lambda_r}(\mu) + \chi_{\Lambda_{r-1}}(\mu) =  E_{r}(z_1^{1/2} + z_1^{-1/2}, \dots , z_r^{1/2} + z_r^{-1/2}) \ . \end{equation}

Suppose that $4|r$, and let $\mu = \vphi$ in \eqref{Dsevlastcharseqn}.  Let $\zeta_j = \exp\left[\pi \ii \frac{\vphi^{j-1}}{\kappa}\right]$.  Then Fact \ref{fact:charge} implies that $\chi_{\Lambda_r}(\vphi) = E_{r}(z_1^{1/2} + z_1^{-1/2}, \dots , z_r^{1/2} + z_r^{-1/2})$, where $z_j^{1/2} + z_j^{-1/2} = \ii (\zeta_j^{-1} - \zeta_j)$, and $z_{r+1-j}^{1/2} + z_{r+1-j}^{-1/2} = (\zeta_j^{-1} + \zeta_j)$, for $1 \leq j \leq \frac{r}{2}$ .  By \eqref{Bcharsym}, $\chi'_{\Lambda'_{\frac{r}{2}}}(\wtild{\vphi}) = \prod_{j=1}^{\frac{r}{2}}({z'_j}^{1/2} + {z'_j}^{-1/2})$, where ${z'_j}^{1/2} + {z'_j}^{-1/2} = \ii(\zeta_j^{-2} - \zeta_j^2)$.  Finally, if $4 \nmid r$, then Equation \eqref{Dsevlastcharseqn} now implies that $\chi_{\Lambda_{r-1}}(\vphi) = E_{r}(z_1^{1/2} + z_1^{-1/2}, \dots , z_r^{1/2} + z_r^{-1/2})$, and so the above argument applies. $\square$


\paragraph{The simple-current $J_vJ_s$, $r$ even.}

Recall the conjugation $C_1$ in \eqref{DChargeConj}; we will denote it by $C$ in this section.  The identity $J_vJ_s = CJ_sC$ implies that $\vphi$ is a fixed point of $J_vJ_s$ if and only if $C\vphi$ is a fixed point for $J_s$.  Therefore, the fixed points for $J_vJ_s$ are all points $\vphi = C\vphi_s$ such that $J_s \vphi_s = \vphi_s$.

By Fact \ref{fact:charge}, $\chi_{\Lambda_n}(\vphi) = 0$ whenever $n$ is odd, or $n = r-1$ ($2 || r$), $n=r$ ($4|r$).  Let $0 \leq m \leq \frac{r-2}{2}$.  Then \begin{equation} \chi_{\Lambda_{2m}}(\vphi) = \chi_{\Lambda_{2m}}(C\vphi_s) = \frac{S_{\Lambda_{2m}, C\vphi_s}}{S_{0, C\vphi_s}} = \frac{S_{C \Lambda_{2m}, \vphi_s}}{S_{C0, \vphi_s}} = \chi_{\Lambda_{2m}}(\vphi_s) \ , \nonumber  \end{equation} where the last equality is because $C$ acts trivially on $\Lambda_{2m}$ and $0$.  Now suppose $2||r$.  Then \begin{equation}  \chi_{\Lambda_r}(\vphi) = \chi_{\Lambda_r}(C\vphi_s) = \frac{S_{\Lambda_r, C\vphi_s}}{S_{0, C\vphi_s}} =\frac{S_{\Lambda_{r-1}, \vphi_s}}{S_{0, \vphi_s}} = \chi_{\Lambda_{r-1}}(\vphi_s) \ ,  \nonumber \end{equation} and similarly, if $4|r$, then $\chi_{\Lambda_{r-1}}(\vphi) = \chi_{\Lambda_r}(\vphi_s)$.

Thus we have the fixed point factorisation:

\begin{fpf}\label{FPF:DreJvJs} Let $\vphi$ be a fixed point for $J_vJ_s$, and let $\wtild{\vphi}$ be as in \eqref{DevenJsphiprime}.  Then \begin{equation}\label{Drevenfpfvseqn} \chi_{\Lambda_{n}}(\vphi) = \left\{ \begin{array}{ll} (-1)^{m} \sum_{\ell = 0}^m \chi'_{\Lambda'_\ell}(\wtild{\vphi}) & \mbox{ if } 0 \leq n=2m \leq r-2 \\ (-1)^{\frac{r}{2}}\chi'_{\Lambda'_{\frac{r}{2}}} & \mbox{ if } n = r-1 \mbox{ and } 4|r \mbox{ , or } n=r \mbox{ and } 2||r \\ 0 & \mbox{otherwise}  \end{array} \right. \  , \end{equation} where primes denote $B_{\frac{r}{2}}^{(1)}$, level $k/2$ quantities. \end{fpf}

\subsection{The algebra $A_{2r-1}^{(2)}$}

Fixed points are of the form \begin{equation}\label{Atwphi} \vphi = (\vphi_1; \vphi_1, \dots , \vphi_r) \ , \end{equation} where \begin{equation} \vphi_1 + \cdots \vphi_r = k/2 \ . \end{equation} Define \begin{equation}\label{Atwphiprime} \wtild{\vphi} = (\vphi_r; \dots , \vphi_1) \ ; \end{equation} this is a level $k/2$, $C_{r-1}^{(1)}$ weight.

The fixed point factorisation is \begin{fpf}\label{fpf:Atw} Let $\vphi$, $\wtild{\vphi}$ be as in equations \eqref{Atwphi}, \eqref{Atwphiprime} respectively.  Then \begin{equation}\label{Atwfpfeqn} \chi_{\Lambda_n}(\vphi) = \left\{ \begin{array}{ll} (-1)^n \left(\chi'_{\Lambda'_n}(\wtild{\vphi}) + \chi'_{\Lambda'_{n-1}}(\wtild{\vphi})\right) & \mbox{ if } 0 \leq n \leq r-1 \\ 0 & \mbox{ if } n=r \end{array} \right. \ , \end{equation} where primes denote $C_{r-1}^{(1)}$ level $k/2$ quantities. \end{fpf}

\noindent \emph{Proof.} Let $\vphi^j = j + \sum_{\ell=1}^j$, with $\vphi^0 = 0$, and $\xi_j := \exp\left[2 \pi \ii \frac{\vphi^j}{\kappa}\right]$.  The orthogonal coordinates are $\vphi^+[j] = \kappa/2 -\vphi^{j-1}$, for $j = 1, \dots , r$.  Fact \ref{fact:charge} implies $\chi_{\Lambda_r}(\vphi) = 0$, and putting $\nu = \vphi$ into \eqref{Atw2charsym}, we find that for $0 \leq n \leq r-1$, \begin{equation}\label{Atwcharsimp} \chi_{\Lambda_n}(\vphi) = (-1)^n (\veps_n + \veps_{n-1} - \veps_{n-2} - \veps_{n-3}) \ , \end{equation} where $\veps_n = E_n(\xi_1, \dots ,\xi_{r-1}, \xi_1^{-1}, \dots ,\xi_{r-1}^{-1})$.
Now consider $C_{r-1}^{(1)}$, level $k/2$.  The orthogonal coordinates of $\wtild{\vphi}$ are $\wtild{\vphi}^+[r-j] = \vphi^{r-j}$ for $j = 1, \dots , r-1$, and putting $\mu = \wtild{\vphi}$ into \eqref{Ccharsym} yields \begin{equation}\label{Atwprimecharsimp} \chi'_{\Lambda'_n}(\wtild{\vphi}) = \veps_n - \veps_{n-2} \ . \end{equation}  Equations \eqref{Atwcharsimp} and \eqref{Atwprimecharsimp} give the fixed point factorisation \eqref{Atwfpfeqn}. $\square$

\subsection{The algebra $D_{r+1}^{(2)}$}

Fixed points of $J$ are of the form \begin{equation}\label{Dtwphi} \vphi = (\vphi_0;  \vphi_1, \dots , \vphi_1, \vphi_0) \ . \end{equation} The fixed point factorisation depends on whether $r$ is odd or even.  In both cases, Fact \ref{fact:charge} implies that $\chi_{\Lambda_n}(\vphi) = 0$ whenever $n$ is odd.

Throughout this section, we will let $\vphi^j = 2j-1 + \vphi_0 + 2 \sum_{\ell=1}^{j-1}$; $\zeta_j = \exp\left[\pi \ii \frac{\vphi^+[j]}{\kappa}\right]$, $\xi = \zeta^2$, and $\veps_m := E_m(\xi_1, \dots ,\xi_{r'}, \xi_1^{-1}, \dots ,\xi_{r'}^{-1})$, where $r'$ is given in each case.

\paragraph{The case $r$ is odd.}  Let $r':= (r-1)/2$.  The weight $\vphi$ satisfies \begin{equation}\label{Dtwphiweightaddup} \vphi_0 + 2(\vphi_1 + \cdots + \vphi_{r'}) = k/2 \ . \end{equation}  Define \begin{equation}\label{Dtwphiprime} \wtild{\vphi} = (\vphi_0; \dots , \vphi_{\frac{r-1}{2}}) \ . \end{equation}  Then $\wtild{\vphi}$ is a $A_{2\left(\frac{r-1}{2}\right)}^{(2)}$, level $k/2$ weight.  We have the fixed point factorisation

\begin{fpf}\label{fpf:Dtw} Let $m \in \{0, \dots , \frac{r-1}{2}\}$, and let $\vphi$, $\wtild{\vphi}$ be as in eqns \eqref{Dtwphi}, \eqref{Dtwphiprime} respectively.  Then \begin{equation}\label{Dtwfpfeqn} \chi_{\Lambda_{n}}(\vphi) = \left\{ \begin{array}{ll} \chi'_{\Lambda'_m}(\wtild{\vphi}) & \mbox{ if } n=2m \\ 0 & \mbox{ if } n \mbox{ is odd } \end{array} \right. , \end{equation} where primes denote $A_{2\left(\frac{r-1}{2}\right)}^{(2)}$ level $k/2$ quantities. \end{fpf}

\noindent \emph{Proof.} Let $n=2m$, where $0 \leq m \leq r'$.  By \eqref{Dtwphiweightaddup}, the orthogonal coordinates of $\vphi$ are: $\vphi^+[j] = \kappa - \vphi^j$ and $\vphi^+[r+1-j] = \vphi^j$ for $1 \leq j \leq r'$, and $\vphi[r'+1] = \kappa/2$.  Thus, putting $\nu = \vphi$ in \eqref{Dtwcharsym} and using Facts \ref{Erecursion} and \ref{lemma:Eeven} yields the expression \begin{equation}\label{Dtwcharsimp} \chi_{\Lambda_n(\vphi)} = (-1)^m (\veps_m - \veps_{m-2}) \ . \end{equation}
Consider $A_{2r'}^{(2)}$ at level $k/2$.  Equation \eqref{Dtwphiweightaddup} gives the orthogonal coordinates of $\wtild{\vphi}$ as $\wtild{\vphi}^+[j] = \kappa/4 - \vphi^j/2$, for $j= 1, \dots , r'$.  Putting $\nu = \wtild{\vphi}$ into \ref{AtwSymChar} yields the expression $\chi'_{\Lambda'_m}(\wtild{\vphi}) = (-1)^m (\veps_m-\veps_{m-2})$.  Comparing with \eqref{Dtwcharsimp} gives the fixed point factorisation \eqref{Dtwfpfeqn}. $\square$

\paragraph{The case $r$ is even.} Let $r'= r/2$.  Fixed points satisfy \begin{equation}\label{Dtwevphiweightaddup} \vphi_0 + 2(\vphi_1 + \cdots + \varphi_{r'-1}) + \vphi_{r'} = k/2 \ . \end{equation}  Define \begin{equation}\label{Dtwevphiprime} \wtild{\vphi} = (\vphi_0; \dots , \vphi_{r'}) \ ; \end{equation} then $\wtild{\vphi}$ is a $D_{\frac{r}{2}+1}^{(2)}$ weight, at level $k/2$.

The fixed point factorisation is \begin{fpf} Let $m \in \{0, \dots , r/2\}$, and let $\vphi$, $\wtild{\vphi}$ be as in eqns \eqref{Dtwphi}, \eqref{Dtwevphiprime} respectively.  Then \begin{equation}\label{fpfeqnDtwev} \chi_{\Lambda_n}(\vphi) = \left\{ \begin{array}{ll} \sum_{i=0}^m (-1)^i \chi'_{\Lambda'_{m-i}}(\wtild{\vphi}) & \mbox{ if } n=2m \\ 0 & \mbox{ otherwise} \end{array} \right. \ , \end{equation} where primes denote $D_{r+1}^{(2)}$, level $k/2$ quantities. \end{fpf}

\noindent \emph{Proof.} Let $n=2m$, where $m \in \{0, \dots , r/2\}$.  By the same argument as above, the orthogonal coordinates become $\vphi^+[j] = \kappa - \varphi^j$ and $\vphi^+[r+1-j] = \vphi^j$, for $1 \leq j \leq r/2$, and the characters evaluated at $\vphi$ are \begin{equation}\label{Dtwevchareqnsimp} \chi_{\Lambda_n}(\vphi) = (-1)^m (\veps_m + \veps_{m-1}) \ . \end{equation}

Putting $\nu = \wtild{\vphi}$ into \eqref{Dtwcharsym} gives the expression \begin{equation}\label{Dtwevprimchareqnsimp} (-1)^m(\veps_m - \veps_{m-2}) \end{equation} for the characters of the fixed point factorisation algebra.  Eqns \eqref{Dtwevchareqnsimp} and \eqref{Dtwevprimchareqnsimp} now yield  $\chi_{\Lambda_{2m}}(\vphi) + \chi_{\Lambda_{2(m-1)}}(\vphi) = \chi'_{\Lambda'_m}(\wtild{\vphi})$, from which \eqref{fpfeqnDtwev} follows. $\square$


\section{Concluding remarks}

This paper, together with \cite{FPF}, establishes that a fixed point factorisation exists for the classical WZW models.  The motivation for this paper was the development of the tool which allows us to calculate NIM-reps, D-brane charges and charge groups, which will follow in Part 2.  However, the existence of this $S$-matrix feature for the WZW models raises some further questions.  Among them:

\begin{itemize}
\item Does fixed point factorisation occur in other contexts?  For example, what would a fixed point factorisation look like for finite group modular data?  Their data looks quite different from the WZW data (an introduction to finite group modular data is given in \cite{FinGroup}).  Ultimately, we would like to know whether fixed point factorisation is a feature of all RCFTs.

\item What is the connection between the fixed point factorisation algebras and the orbit Lie algebras?

\item What is a conceptual explanation for fixed point factorisation?
\end{itemize}

As we mentioned in the introduction, fixed points can be a source of difficulty, and it is our hope that, beyond the applications explored in Part 2 of this paper, fixed point factorisation will be used as a tool elsewhere in mathematics and physics.


\noindent {\bf Acknowledgements}
We are grateful to Terry Gannon for helpful discussions during this research and writing of the paper.


\end{document}